\documentclass[12pt]{article}

\usepackage{amssymb}
\usepackage{amsmath}
\usepackage{amscd}
\usepackage{latexsym}

\usepackage{graphicx}
\usepackage{subfig}

\usepackage{url}

\usepackage{cite}
\usepackage{caption}

\newcommand{\be}{\begin{equation}}
\newcommand{\ee}{\end{equation}}
\newcommand{\Dlt}{\Delta}
\newcommand{\dlt}{\delta}
\newcommand{\prt}{\partial}
\newcommand{\br}{{\bf r}}
\newcommand{\bk}{{\bf k}}
\newcommand{\bfe}{{\bf e}}
\newcommand{\bd}{{\bf d}}
\newcommand{\bP}{{\bf P}}
\newcommand{\bn}{{\bf n}}
\newcommand{\bt}{\beta}
\newcommand{\vp}{\varphi}
\newcommand{\ep}{\varepsilon}
\newcommand{\al}{\alpha}
\newcommand{\ra}{\rightarrow}
\newcommand{\sgm}{\sigma}

\newcommand{\gm}{\gamma}
\newcommand{\om}{\omega}
\newcommand{\Om}{\Omega}

\newcommand{\dgr}{\dagger}
\newcommand{\lbd}{\lambda}
\newcommand{\Lbd}{\Lambda}

\newcommand{\rgl}{\rangle}
\newcommand{\lgl}{\langle}

\begin{document}

\begin{center}

{\Large{\bf Bose-condensed atomic systems with nonlocal interaction potentials} \\ [5mm]

V.I. Yukalov$^{1,*}$ and E.P. Yukalova$^2$ } \\ [3mm]

{\it
$^1$Bogolubov Laboratory of Theoretical Physics, \\
Joint Institute for Nuclear Research, Dubna 141980, Russia \\ [3mm]

$^2$Laboratory of Information Technologies, \\
Joint Institute for Nuclear Research, Dubna 141980, Russia }

\end{center}

\vskip 5cm

\begin{abstract}

The general approach for describing systems with Bose-Einstein condensate,
where atoms interact through nonlocal pair potentials, is presented. A 
special attention is paid to nonintegrable potentials, such as the dipolar
interaction potential. The potentials that are not absolutely integrable
can have not well defined Fourier transforms. Using formally such not 
defined Fourier transforms leads to unphysical conclusions. For making 
the Fourier transform well defined, the interaction potential has to be 
regularized. This is illustrated by the example of dipolar interactions.   

\end{abstract}

\vskip 1cm

{\parindent = 0pt
{Keywords}: Bose-Einstein condensate, nonlocal interaction potentials,
dipolar interactions, regularization, screening, spectrum anisotropy

\vskip 2cm

{\bf $^*$corresponding author}: V.I. Yukalov

{\bf E-mail}: yukalov@theor.jinr.ru   }

\newpage

\section{Introduction}

Atomic systems exhibiting Bose-Einstein condensation are widely studied
both theoretically and experimentally, as can be inferred from the books 
\cite{Lieb_1,Letokhov_2,Pethick_3} and review articles 
\cite{Courteille_4,Andersen_5,Yukalov_6,Bongs_7,Yukalov_8,Posazhennikova_9,
Proukakis_10,Yurovsky_11,Yukalov_12,Yukalov_13}. Trapped dilute atomic gases
are often characterized by local interaction potentials of the delta-function 
type. But recently, trapped atoms interacting through dipolar forces have been
condensed (see reviews \cite{Griesmaier_14,Baranov_15,Baranov_16}). Dipolar 
interactions are nonlocal and long-range and their theoretical description is 
more complicated than that of locally interacting atoms, because of which one
usually considers only the Bogolubov approximation that is valid at temperatures 
close to zero and asymptotically weak interactions.  

Moreover, the dipolar interaction potential does not possesses a well defined
Fourier transform. Formally taking this transform and using it leads to 
unphysical consequences.

The aim of the present paper is to suggest a self-consistent approach for 
atomic systems with nonlocal interaction potentials, valid for finite 
temperatures and for interactions of any strength. A special attention is
paid to nonintegrable potentials, whose Fourier transforms are not well 
defined. As a typical example of this kind, dipolar interactions are treated. 
We show that such interactions need to be regularized in order to get a 
correct description of atomic systems. Such a regularization is necessary 
for atoms of any statistics, whether Bose or Fermi. For concreteness, we 
consider here atomic systems of Bose-Einstein statistics. 

Throughout the paper, we use the notation where the Planck and Boltzmann 
constants are set to unity, $ \hbar=1$, $k_B=1$.

\section{Self-consistent approach}

A self-consistent approach for describing Bose-condensed systems has been
developed in Refs. \cite{Yukalov_17,Yukalov_18,Yukalov_19}. This approach
resolves the Hohenberg-Martin dilemma \cite{Hohenberg_20} of conserving 
versus gapless theories and provides a theory that is conserving as well
as gapless. 

The general form of the energy Hamiltonian is
\be
\label{1}
 \hat H = \int \hat\psi^\dgr(\br) \left ( - \; \frac{\nabla^2}{2m}
+ U \right )   \hat\psi(\br) \; d\br \;  + \; 
\frac{1}{2} \int \hat\psi^\dgr(\br) \hat\psi^\dgr(\br') \Phi(\br-\br')
\hat\psi(\br') \hat\psi(\br) \; d\br d\br' \; ,
\ee
where the field operators satisfy the Bose commutation relations, 
$U = U({\bf r})$ is an external potential, if any, and $\Phi ({\bf r})$
is a nonlocal interaction potential. The field operators depend on time 
$t$, which, for simplicity, is not explicitly shown. 

For the occurrence of Bose-Einstein condensation, the necessary and 
sufficient condition is the global gauge symmetry breaking
\cite{Lieb_1,Yukalov_12,Yukalov_13,Yukalov_21}. A convenient way of 
breaking the gauge symmetry is by employing the Bogolubov shift
\cite{Bogolubov_22,Bogolubov_23} representing the field operators as
\be
\label{2}
 \hat\psi(\br) = \eta(\br) + \psi_1(\br) \;  ,
\ee
where $\eta ({\bf r})$ is the condensate function and $\psi_1({\bf r})$
is the operator of uncondensed atoms.

Note that the Bogolubov shift is an exact canonical transformation, but
not an approximation, as one sometimes writes. 

To avoid double counting, the condensate function and the field operator
of uncondensed atoms are assumed to be orthogonal, 
\be
\label{3}
 \int \eta^*(\br) \psi_1(\br)\; d\br = 0 \;  .
\ee

The condensate function plays the role of the functional order parameter,
such that
\be
\label{4}
  \eta(\br) = \lgl \hat \psi(\br) \rgl \;.
\ee
The latter implies that
\be
\label{5}
  \lgl \psi_1(\br) \rgl = 0 \;.
\ee

The condensate function is normalized to the number of condensed atoms
\be
\label{6}
 N_0 = \int | \eta(\br) |^2 \; d\br \;  .
\ee
While the number of uncondensed atoms is the statistical average 
\be
\label{7}   
N_1 = \lgl \hat N_1 \rgl
\ee
of the operator
$$
 \hat N_1 \equiv \int \psi^\dgr_1(\br)\psi_1(\br) \; d\br \;  .
$$

The evolution equation for the condensate function can be written as
\be
\label{8}
 i \; \frac{\prt}{\prt t} \; \eta(\br,t) = 
\left \lgl \frac{\dlt H}{\dlt\eta^*(\br,t)} \right \rgl \;  .
\ee
And the equation of motion for the operator of uncondensed atoms reads as
\be
\label{9}
 i \; \frac{\prt}{\prt t} \; \psi_1(\br,t) = 
\frac{\dlt H}{\dlt\psi_1^\dgr(\br,t)} \; .
\ee
The grand Hamiltonian generating these equations is
\be
\label{10}
 H = \hat H - \mu_0 N_0 - \mu_1 \hat N_1 - \hat\Lbd \;  ,
\ee
where the first term is the energy operator (\ref{1}), the Lagrange 
multipliers $\mu_0$ and $\mu_1$ guarantee the validity of the normalization
conditions (\ref{6}) and (\ref{7}), while the last term, having the form 
$$
 \hat\Lbd = \int \left [ \lbd(\br) \psi_1^\dgr(\br) +
 \lbd^*(\br) \psi_1(\br) \right ] \; d\br \; ,
$$
preserves the validity of condition (\ref{5}). It has been proved 
\cite{Yukalov_13,Yukalov_24} that the variational equation (\ref{9}) 
is equivalent to the Heisenberg equation of motion.  
 
In systems with broken gauge symmetry, in addition to the single-particle
density matrix 
\be
\label{11}
 \rho_1(\br,\br') = \lgl \psi_1^\dgr(\br') \psi_1(\br) \rgl \;  ,
\ee
there exists the anomalous matrix
\be
\label{12}
 \sgm_1(\br,\br') = \lgl \psi_1(\br') \psi_1(\br) \rgl \;   .
\ee

The condensate function defines the density of condensed atoms
\be
\label{13}
  \rho_0(\br) = | \eta(\br) |^2 \; .
\ee
The diagonal elements of the single-particle density matrix give
the density of uncondensed atoms

\be
\label{14}
 \rho_1(\br) = \rho_1(\br,\br) =  
\lgl \psi_1^\dgr(\br) \psi_1(\br) \rgl \; .
\ee
And the diagonal elements of the anomalous matrix define the density
$$
|\sgm_1(\br) | = | \lgl \psi_1(\br) \psi_1(\br) \rgl |
$$
of pair-correlated atoms. The total atomic density is the sum
\be
\label{15}
   \rho(\br) = \rho_0(\br) + \rho_1(\br) 
\ee
yielding the total number of atoms
\be
\label{16}
 N = \int \rho(\br) \; d\br = N_0 + N_1 \;  .
\ee
The partial atomic ratios give the fractions of condensed, $n_0$, and 
uncondensed, $n_1$ atoms, respectively,
\be
\label{17}
 n_0 = \frac{N_0}{N} \; , \qquad n_1 = \frac{N_1}{N} \qquad
( n_0 + n_1 = 1 ) \;  .
\ee
   
For an equilibrium system, the statistical operator has the form
\be
\label{18}
 \hat\rho = \frac{1}{Z} \; e^{-\bt H} \qquad
\left ( Z = {\rm Tr} e^{-\bt H} \right ) \;  ,
\ee
with $\beta = 1/T$ being the inverse temperature, and with the same grand 
Hamiltonian (\ref{10}). 

The superfluid fraction is given by the expression
\be
\label{19}
 n_s = 1 - \; \frac{{\rm var}(\hat\bP)}{m T N d} \;  ,
\ee
where $d$ is the real-space dimensionality, the operator of momentum is
$$
 \hat P = 
\int \hat\psi^\dgr(\br) ( - i \vec{\nabla} )\hat \psi(\br) \; d\br \; ,
$$
and the variance of an arbitrary operator $\hat{A}$ is defined as
$$
 {\rm var}(\hat A) \equiv \frac{1}{2} \lgl \hat A^{+} \hat A +
\hat A \hat A^{+} \rgl - | \lgl \hat A \rgl |^2 \;  .
$$
For a self-adjoint operator, this yields
$$
 {\rm var}(\hat A) =  \lgl \hat A^2 \rgl - \lgl \hat A \rgl^2 \;  .
$$
In equilibrium, the average momentum is zero, 
$$
  \lgl \hat\bP \rgl = 0 \; ,
$$
which, for dimensionality $d = 3$, leads to
\be
\label{20}
 n_s =  1 -  \; \frac{\lgl \hat\bP^2\rgl}{3 m T N} \;  .
\ee

\section{Uniform systems}

When there is no external potential, $U = 0$, or when the trap is 
sufficiently large, the system can be treated as uniform. Then it 
is possible to Fourier transform the field operators,
\be
\label{21}
 \psi_1(\br) = 
\frac{1}{\sqrt{V}} \; \sum_{k\neq 0} a_k e^{i\bk\cdot\br} \; , 
\qquad
 a_k = 
\frac{1}{\sqrt{V}} \; \int \psi_1(\br) e^{- i\bk\cdot\br} d\br  \;  .
\ee
Usually, one assumes that the interaction potential also enjoys the
Fourier transformation
\be
\label{22}
\Phi(\br) = \frac{1}{V} \; \sum_{k} \Phi_k e^{i\bk\cdot\br} \; , 
\qquad
 \Phi_k =  \int \Phi(\br) e^{- i\bk\cdot\br} d\br \;    .
\ee
If the Fourier transform is well defined, there exists the limit
\be
\label{23}
 \Phi_0 =\lim_{k\ra 0} \Phi_k = \int \Phi(\br) \; d\br \;  .
\ee

Note that if the trapping potential $U$ is not zero, but sufficiently 
smooth, it is possible to resort to the local-density approximation, for
which the similar Fourier transforms are also assumed.

The single-particle density matrix becomes
\be
\label{24}
 \rho_1(\br,\br') =  \frac{1}{V} \; \sum_{k\neq 0} n_k e^{i\bk\cdot(\br-\br')} \;  ,
\ee
with the momentum distribution
\be
\label{25}
 n_k = \lgl a_k^{+} a_k \rgl \;  .
\ee
And the anomalous average is
\be
\label{26}
 \sgm_1(\br,\br') =  \frac{1}{V} \; \sum_{k\neq 0} \sgm_k e^{i\bk\cdot(\br-\br')} \;  ,
\ee
with 
\be
\label{27}
 \sgm_k = \lgl a_k a_{-k} \rgl \;  .
\ee
 
The condensate function and atomic densities do not depend in the spatial
variable. The total average density reads as
\be
\label{28}
 \rho = \rho_0 + \rho_1 \;  ,
\ee
with the condensate density
\be
\label{29}
\rho_0 = |\eta|^2
\ee
and the density of uncondensed atoms
\be
\label{30}
 \rho_1 = \frac{1}{V} \sum_{k\neq 0} n_k \;  .
\ee
The diagonal anomalous average becomes
\be
\label{31}
 \sgm_1 = \frac{1}{V} \sum_{k\neq 0} \sgm_k \;  .
\ee
   
In the Hartree-Fock-Bogolubov approximation, we obtain 
\cite{Yukalov_12,Yukalov_13} the momentum distribution
\be
\label{32}
n_k = \frac{\om_k}{2\ep_k}\; 
\coth \left ( \frac{\ep_k}{2T} \right ) - \; \frac{1}{2} \;   ,
\ee
and the anomalous average (\ref{27}) reads as
\be
\label{33}
 \sgm_k = -\; \frac{\Dlt_k}{2\ep_k}\; 
\coth \left ( \frac{\ep_k}{2T} \right )  \;  .
\ee
Here we use the notations
\be
\label{34}
\om_k = \frac{k^2}{2m} + \rho \Phi_0 + \rho_0 \Phi_k +
\frac{1}{V} \sum_{p\neq 0} n_p \Phi_{k+p} \; -  \; \mu_1
\ee
and
\be
\label{35}
\Dlt_k = \rho_0 \Phi_k + \frac{1}{V} \sum_{p\neq 0} \sgm_p \Phi_{k+p}
\ee
defining the spectrum of collective excitations
\be
\label{36}
 \ep_k = \sqrt{\om_k^2-\Dlt_k^2} \;  .
\ee

In equilibrium, 
\be
\label{37}
 i\; \frac{\prt\eta}{\prt t} = 
\left \lgl \frac{\dlt H}{\dlt\eta^*} \right \rgl = 0 \;  ,
\ee
which is equivalent to the variational condition
\be
\label{38}
 \left \lgl \frac{\dlt H}{\dlt\rho_0} \right \rgl = 0 \; .
\ee
From the latter, we get the condensate chemical potential  
\be
\label{39} 
 \mu_0 = \rho \Phi_0 + 
\frac{1}{V} \sum_{k\neq 0} ( n_k + \sgm_k) \Phi_k \;  .
\ee

The condition of the condensate existence \cite{Yukalov_12,Yukalov_13}
\be
\label{40}
 \lim_{k\ra 0} \; \frac{1}{n_k} = 0   
\ee
requires a gapless spectrum, such that
\be
\label{41}
 \lim_{k\ra 0} \ep_k = 0 \; ,
\ee
in agreement with the Bogolubov \cite{Bogolubov_22,Bogolubov_23} and 
Hugenholtz-Pines \cite{Hugenholtz_25} theorems. Condition (\ref{41})
yields the chemical potential of uncondensed atoms
\be
\label{42}
 \mu_1 = \rho \Phi_0 + \frac{1}{V} \sum_{k\neq 0} ( n_k - \sgm_k) \Phi_k \;  .
\ee
Then expression (\ref{34}) takes the form
\be
\label{43}
 \om_k = \frac{k^2}{2m} + \rho_0 \Phi_k +  
\frac{1}{V} \sum_{p\neq 0} ( n_p \Phi_{k+p} - n_p \Phi_p + \sgm_p \Phi_p ) \;.
\ee
In that way, for the spectrum of collective excitations (\ref{36}) we find
$$
\ep_k^2 = \left [ \frac{k^2}{2m} +   
\frac{1}{V} \sum_{p\neq 0} ( n_p - \sgm_p) (\Phi_{k+p} - \Phi_p ) \right ] \times
$$
\be
\label{44}
 \times
\left \{  \frac{k^2}{2m} + 2\rho_0 \Phi_k + \frac{1}{V} \sum_{p\neq 0} 
 [ \; ( n_p + \sgm_p) \Phi_{k+p} - (n_p - \sgm_p) \Phi_p \; ] \right \} \; .
\ee
 
In order to simplify the formulas, we notice that expressions (\ref{32}) and
(\ref{33}) strongly increase as $k \ra 0$. Therefore the main contribution 
in the summations, containing $n_p$ and $\sigma_p$, comes from the region 
of small $p$. This suggests the possibility of using the approximations
\be
\label{45}
 \sum_{p\neq 0} n_p \Phi_{k+p} \cong \Phi_k \sum_{p\neq 0} n_p \; , \qquad
\sum_{p\neq 0} \sgm_p \Phi_{k+p} \cong \Phi_k \sum_{p\neq 0} \sgm_p \;   .
\ee
Then the chemical potentials (\ref{39}) and (\ref{42}) become
\be
\label{46}
\mu_0 = (\rho +\rho_1 + \sgm_1 ) \Phi_0
\ee
and, respectively,
\be
\label{47}
 \mu_1 = (\rho +\rho_1 - \sgm_1 ) \Phi_0 \; .
\ee
Equation (\ref{43}) reduces to
\be
\label{48}
\om_k = \frac{k^2}{2m} + \rho \Phi_k - (\rho_1 - \sgm_1) \Phi_0 \;   ,
\ee
while expression (\ref{35}) becomes
\be
\label{49}
 \Dlt_k = (\rho_1 + \sgm_1) \Phi_k \;  .
\ee
The collective spectrum (\ref{44}) takes the form
\be
\label{50}
 \ep_k^2 = \left [ \frac{k^2}{2m} + (\rho_1 - \sgm_1) ( \Phi_k - \Phi_0 ) \right ]
\left [ \frac{k^2}{2m} + (\rho +\rho_0 + \sgm_1)\Phi_k - 
(\rho_1-\sgm_1) \Phi_0  \right ] \;  .
\ee

\section{Thermodynamic characteristics}

Having spectrum (\ref{50}), it is possible to calculate the grand potential
\be
\label{51}
 \Om = E_B + T \sum_k \ln \left ( 1 - e^{-\bt\ep_k} \right ) \;  ,
\ee
where
$$
E_B = - \; \frac{1}{2} \; N\rho \Phi_0 - 
\rho_0 \sum_p ( n_p + \sgm_p ) \Phi_p \; -
$$
\be
\label{52}
-  \; \frac{1}{2V} \sum_{kp} ( n_k n_p + \sgm_k\sgm_p ) \Phi_{k+p} +
\frac{1}{2} \sum_k (\ep_k - \om_k) \; .
\ee
With approximation (\ref{45}), we have
\be
\label{53}
 E_B =  -\; \frac{1}{2} V\Phi_0 \left [ \rho^2 + 2\rho_0 (\rho_1 + \sgm_1 )
+ \rho_1^2 + \sgm_1^2 \right ] + \frac{1}{2} \sum_k (\ep_k - \om_k) \;  .
\ee
Using here the fractions of condensed and uncondensed atoms
\be
\label{54}
 n_0 = \frac{\rho_0}{\rho} \; , \qquad    n_1 =  \frac{\rho_1}{\rho}
\ee
and the notation
\be
\label{55}
 \sgm =  \frac{\sgm_1}{\rho}  ,
\ee
we get
\be
\label{56}
  E_B =  -\; \frac{1}{2} N\rho\Phi_0 \left [ 1 + 2n_0 (n_1 + \sgm )
+ n_1^2 + \sgm^2 \right ] + \frac{1}{2} \sum_k (\ep_k - \om_k) \;  .
\ee

The average of the grand Hamiltonian can be represented as
\be
\label{57}
\lgl H \rgl = \lgl \hat H \rgl - \mu N \; ,
\ee
which defines the system chemical potential
\be
\label{58}
 \mu = \mu_0 n_0 + \mu_1 n_1 \;  .
\ee
For the latter, we find
\be
\label{59}
\mu = \rho \Phi_0 + 
\frac{1}{V} \sum_{k\neq 0} [\; n_k + (n_0 - n_1 ) \sgm_k \; ] \Phi_k  ,
\ee
which, in approximation (\ref{45}), reads as
\be
\label{60}
 \mu = \rho \Phi_0 [\; 1 + n_1( 1 - \sgm ) + n_0 \sgm \; ] \; .
\ee
With the given grand potential, it is straightforward to calculate any required
thermodynamic characteristics. 

It is necessary to notice that all thermodynamic characteristics depend on the 
quantity $\Phi_0$ that, hence, has to be well defined. Even in the simplest
Bogolubov approximation, when 
$$
 \frac{\rho_1}{\rho_0} \ll 1 \; , \qquad \frac{|\sgm_1|}{\rho_0} \ll 1 \; ,
$$
so that it is admissible to neglect $\rho_1$ and $\sigma_1$, as compared to
the condensate density $\rho_0 \ra \rho$, we have
$$
 \mu = \mu_0 = \mu_1 = \rho \Phi_0  
$$
and 
$$
E_B = -\; \frac{1}{2} \; N \rho \Phi_0 + 
\frac{1}{2} \sum_k ( \ep_k - \om_k) \;   .
$$
Anyway, thermodynamic characteristics do depend on $\Phi_0$. 

Thus the limit $\Phi_0$ has to be well defined. It must be a scalar
in order that thermodynamic characteristics be scalar quantities. This
implies that the Fourier transform of the interaction potential has to 
be correctly defined, giving an unambiguous limit (\ref{23}).

\section{Fourier transform}

It is useful to remember the conditions when Fourier transforms can be 
correctly defined, since not each function enjoys a well defined Fourier 
transform. For this purpose, let us remind some known mathematical facts.
For simplicity, we recall these facts for the case of one variable. The 
generalization to several variables is straightforward.  

\vskip 2mm

{\bf Definition of bounded variation}. A function $f(x)$, with 
$x \in (-\infty, \infty)$, is of bounded variation in a finite interval
if in that interval it has only a finite number of extrema and a finite 
number of finite discontinuities.
 
\vskip 2mm

{\bf Dirichlet theorem}. If a function $f(x)$ is of bounded variation 
in any finite interval and is absolutely integrable, such that
$$
 \int_{-\infty}^\infty | f(x) | \; dx < \infty \;  ,
$$
then its Fourier transform
$$
F(k) = \int_{-\infty}^\infty  f(x)  e^{-ikx} \; dx
$$
exists and the inverse Fourier transform gives
$$
 \int_{-\infty}^\infty F(k)  e^{ikx} \; \frac{dk}{2\pi} = 
\frac{1}{2}  [\; f( x - 0 ) + f( x +0 ) \; ] \;.
$$
The details can be found in Ref. \cite{Champeney_26}.

\vskip 2mm
 
For the case of several variables, the sufficient condition for the 
existence of Fourier transform of a potential $\Phi({\bf r})$ is its 
absolute integrability:
\be
\label{61}
 \int | \Phi(\br) | \; d\br < \infty \;  .
\ee
We also may notice that the correct definition of the limit (\ref{23})
requires the validity of interchanging the limit and integration.

\vskip 2mm

{\bf Lebesgue theorem}. A sufficient condition for the interchange of 
a limiting procedure and integration is the absolute integrability of the
considered function. 

\vskip 2mm
For the case of the interaction potential, the limit (\ref{23}) exists, 
so that the change of the orders
$$
\lim_{k\ra 0} \int \Phi(\br) e^{-i\bk \cdot\br} \; d\br =
\int \lim_{k\ra 0} \Phi(\br) e^{-i\bk \cdot\br} \; d\br
$$
is valid, provided that the interaction potential is absolutely integrable,
according to condition (\ref{61}). 

A good illustration of nonlocal potentials is the potential of dipole-dipole 
interactions. Dipoles can be electric, with the dipolar moment ${\bf d}_i$,
or magnetic, with the magnetic moment $\vec{\mu}_i$, where the index $i$
enumerates particles. We shall consider dipoles with the moment ${\bf d}_i$, 
keeping in mind that the same problems concern dipoles with the moment 
$\vec{\mu}_i$. Dipolar interactions are widespread for different kinds of 
condensed matter \cite{Grosso_27}, for polymers \cite{Barford_28}, biological
systems \cite{Cameretti_29,Waigh_30}, and for many magnetic nanoclusters and 
nanomolecules \cite{Yukalov_31,Yukalov_32}.     
    
Two dipoles at the distance $r$ from each other interact through the dipolar
potential
\be
\label{62}
D(\br) = \frac{1}{r^3} \left [ \; ( \bd_1 \cdot \bd_2 ) - 
3 ( \bd_1 \cdot \bn ) ( \bd_2 \cdot \bn ) \; \right ] \; ,
\ee
where
$$
 r \equiv | \br | \; , \qquad \bn \equiv \frac{\br}{r} \; , \qquad
\br \equiv \br_1 - \br _2 \;  .
$$
One often considers polarized dipoles, all directed along the same unit vector
${\bf e}_d$, so that
$$
 \bd_i = d_0 \bfe_d \qquad (d_0 \equiv | \bd_i |) \;  .
$$
Then the dipolar potential reduces to
\be
\label{63}
 D(\br) = \frac{d_0^2}{r^3} \; \left (1 - \cos^2\vartheta \right ) \;  ,
\ee
where $\vartheta$ is the angle between the direction of a dipole and the spatial
vector ${\bf n} = {\bf r}/r$,
\be
\label{64}
 \cos\vartheta  = \left ( \bfe_d \cdot \bn \right ) \; .
\ee
If the dipoles are directed along the axis $z$, then 
$\cos \vartheta = z/r$. 

One formally defines the Fourier transform
\be
\label{65}
 D_k = \int D(\br) e^{-i\bk\cdot\br} \; d\br \;  .
\ee
Then in the general case (\ref{62}), one gets 
\be
\label{66}
 D_k = \frac{4\pi}{3} \left [ 
\frac{3(\bd_1\cdot\bk)(\bd_2\cdot\bk)}{k^2} \; - \; 
(\bd_1\cdot\bd_2) \right ] \;  ,
\ee
with $k = |{\bf k}|$. And for the case of polarized dipoles, one has
\be
\label{67}
 D_k = \frac{4\pi}{3} \; d_0^2 \left ( 3\cos^2\vartheta_k - 1 \right ) \;  ,
\ee
with $\vartheta_k$ being the angle between the dipole direction and the 
vector ${\bf k}$,
\be
\label{68}
 \cos\vartheta_k = \frac{(\bk\cdot\bfe_d)}{k} \;  .
\ee
If the dipole direction is along the axis $z$, then $\vartheta_k = k_z/k$. 

However, there are problems with the dipolar interaction potential. Thus,
it is easy to see that the integral
\be
\label{69}
D_0 \equiv \int D(\br) \; d\br 
\ee
is not well defined. For an infinite system, it has to be understood as an 
improper integral
$$
\int D(\br) \; d\br = \lim_{V\ra\infty} \int_V D(\br) \; d\br  \; .
$$
But if one integrates, first, over spherical angles, one gets zero. While, 
if one integrates, first, over the radius, one gets infinity. Hence
\be
\label{70}
 \lim_{k\ra 0} D_k \neq D_0 \; ,
\ee
contrary to condition (\ref{23}). 

Moreover, the limit $k \ra 0$ for $D_k$ is not defined at all, as is seen 
from the above forms of $D_k$. Such a limit explicitly depends on the angle
(\ref{68}). Usually, one alleges that this anisotropy is appropriate for
a system with dipolar interactions. However, this is absolutely wrong. 
The quantity $D_0$, similar to $\Phi_0$, enters many thermodynamic 
characteristics, as is explained in Sec. 4. The thermodynamic quantities,
such as chemical potential, energy, grand potential, by their meaning 
are scalars and principally cannot be anisotropic. Since the limit $D_0$
is not defined, then all thermodynamic characteristics are not defined,
which has no sense.    

This problem arises because the dipolar potential (\ref{62}) or (\ref{63})
is not absolutely integrable. Really, since
$$
 \int_0^\pi | 1 - 3\cos^2\vartheta | \sin\vartheta \; d\vartheta =
\frac{8}{3\sqrt{3}} \;  ,
$$
we have 
$$
 \int | D(\br) | \; d\br = \lim_{R\ra\infty} \lim_{b\ra 0} \; 
\frac{16\pi}{3\sqrt{3}} \; d_0^2 \; \ln \frac{R}{b} ~ \ra ~ \infty \;  .
$$
This integral tends to infinity for any of the limits, either $R \ra \infty$,
or $b \ra 0$. Therefore the Fourier transform for an absolutely nonintegrable
function may not exist, which is the case for the dipolar potential whose 
Fourier transform is defined neither for $k \ra 0$ nor for $k \ra \infty$.  
The problem arises not due to the anisotropy of the dipolar potential, but due 
to its nonintegrability, which makes the formally introduced Fourier transform 
senseless.

\section{Cutoff regularization}

Actually, the difficulties with the dipolar interaction potential 
are well known in physics of condensed matter \cite{Grosso_27} and 
have been discussed in many publications. Thus the divergence of 
the dipolar potential at short distance, resulting in infinite molecular
polarizability, has been called the {\it polarization catastrophe}
\cite{Applequist_33,Thole_34}. The way out of this catastroph is 
physically transparent, requiring to consider dipolar particles not
as point-like objects, but as finite-size particles described by 
spatial distributions. This way leads to the smearing of dipolar 
interactions at short distances, which can be characterized by 
different smearing functions 
\cite{Burnham_35,Masia_36,Langlet_37,Kanjilal_38,Tarasov_39,Ustunel_40}.
The simplest is the short-range cutoff regularization removing the
$1/r^3$ singularity. This implies the use of the short-range regularized 
potential
\be
\label{71}
D(\br,b) = \Theta(r-b) D(\br) \;  ,
\ee
where $\Theta(r)$ is a unit-step function and $b$ is the sum of two 
atomic radii. In the case of identical atoms, $b$ is the effective atomic 
diameter.   
 
For the Fourier transform 
$$
D_k(b) = \int D(\br,b) e^{-i\bk\cdot \br}\; d\br
$$
of potential (\ref{71}), we have
$$
 D_k(b) = \int_b^\infty r^2 \; dr \; \int_0^\pi \sin\vartheta \; d\vartheta \;
\int_0^{2\pi} d\vp \; D(\br) e^{-i\bk\cdot \br} \; .
$$
Considering the polarized potential (\ref{63}), directing the axis $z$ along the wave 
vector ${\bf k}$, integrating out the angle $\varphi$, and using the notation
$$
x \equiv \cos\vartheta = \frac{\bk\cdot\br}{kr} = \frac{\bk\cdot\bn}{k} \;  ,
$$
we get
$$
D_k(b) = \frac{3}{4} \; D_k \int_b^\infty \frac{dr}{r} \; 
\int_{-1}^1 dx \; \left ( 1 - 3x^2 \right )  e^{-ikrx} \;  .
$$
Integrating over $x$ gives
\be
\label{72}
 D_k(b) = D_k I_k(b) \;  ,
\ee
with 
$$
I_k(b) \equiv 9 \int_{kb}^\infty \left ( \frac{\sin y}{y^4} \; - \; 
\frac{\cos y}{y^3} \; - \; \frac{\sin y}{3y^2} \right ) \; dy \;  ,
$$
where $y = kr$. Taking the last integral results in 
\be
\label{73}
 I_k(b) = \frac{3}{(kb)^3} \left [ \; \sin(kb) - kb \cos (kb) \; \right ] \; .
\ee

Integral (\ref{73}) has the property
$$   
 \lim_{k\ra 0} I_k(b) = \lim_{b\ra 0} I_k(b) =  1 \; .
$$
The presence of the short-range cutoff makes it possible to define the short-wave limit,
that is, the large-$k$ limit,
\be
\label{74}
D_k(b) \simeq - 3D_k\; \frac{\cos(kb)}{(kb)^2} \qquad ( k \ra \infty ) \;   .
\ee
Recall that without the short-range cutoff regularization the large-$k$ limit is not 
defined, as is clear from equation (\ref{67}). Respectively, in the limiting expression
(\ref{74}), it is impossible to set $b$ to zero. 

However, the long-wave limit, when $k \ra 0$, is not defined, since
\be
\label{75}
 D_k(b) \simeq D_k \qquad ( k \ra 0) \; ,
\ee
and we return to the problem discussed in the previous section. This is connected with 
the fact that potential (\ref{71}) is not absolutely integrable, being divergent at 
large $r$.

\section{Screening regularization}

When one considers separate molecules or clusters, such finite small systems require 
only short-range regularization. But for large systems, a long-range regularization, 
taking into account long-range correlations, is also necessary. The necessity of
regularizing the dipolar potential by long-range screening was emphasized by 
Jonscher \cite{Jonscher_41,Jonscher_42,Jonscher_43}. Such a screening is usually described
by an exponential function 
\cite{Tarasov_39,Jonscher_41,Jonscher_42,Jonscher_43,Youjian_44,Baul_45}. 

The dipolar potential that is regularized both for short-range as well as for long-range
interactions can be written in the form
\be
\label{76}
 D(\br,b,\kappa) = \Theta(r-b) D(\br) e^{-\kappa r} \;  ,
\ee
where $\kappa$ is a screening wave vector, hence $1/\kappa$ is a screening radius. This 
potential is absolutely integrable, so that it enjoys a well defined Fourier transform
$$
D_k(b,\kappa) = \int D(\br,b,\kappa) e^{-i\bk\cdot\br} \; d\br \;   .
$$
For the case of polarized dipoles, we have
\be
\label{77}
 D_k(b,\kappa) = D_k I_k(b,\kappa) \;  ,
\ee
with the integral 
\be
\label{78}
I_k(b,\kappa) \equiv 9 \int_{kb}^\infty  \left ( \frac{\sin y}{y^4} \; - \; 
\frac{\cos y}{y^3} \; - \; \frac{\sin y}{3y^2} \right ) e^{-\kappa y/k} \; dy \;  ,
\ee
in which $y = kr$. With the change of the variable $x = r/b$, giving $y = kbx$, we get
\be
\label{79}
 I_k(b,\kappa) = 9 kb \int_1^\infty \left [ \frac{\sin(kbx)}{(kbx)^4} \; - \; 
\frac{\cos(kbx)}{(kbx)^3} \; - \; \frac{\sin (kbx)}{3(kbx)^2} \right ] \; 
e^{-\kappa b x} \; dx \; .
\ee

Removing the screening returns us back to the case of the previous section,
$$
 \lim_{\kappa\ra 0} I_k(b,\kappa) = I_k(b) \; ,
$$
with a not well defined Fourier transform. It is clear that the limits $b\ra 0$ and 
$\kappa \ra 0$ do not commute with the limit $k \ra 0$,
$$
\lim_{b\ra 0} \lim_{\kappa\ra 0} I_k(b,\kappa) =  1 \qquad ( k > 0 ) \; ,
$$
\be
\label{80}
\lim_{k\ra 0} I_k(b,\kappa) = 0 \qquad ( b>0 , \; \kappa > 0 ) \; .
\ee

It is possible to notice that integral (\ref{79}) depends, actually, on two variables
\be
\label{81}
 q \equiv kb \; , \qquad c \equiv \kappa b \;  .
\ee
Therefore integral (\ref{79}) can be represented as
\be
\label{82}
I_k(b,\kappa) = J_q(c)  \;   ,
\ee
with the integral
\be
\label{83}
 J_q(c) = 9 q \int_1^\infty \left [ \frac{\sin(qx)}{(qx)^4} \; - \; 
\frac{\cos(qx)}{(qx)^3} \; - \; \frac{\sin (qx)}{3(qx)^2} \right ] \; 
e^{-c x} \; dx  \;  .
\ee
 
The latter integral can be expressed through the exponential integral function
$$
{\rm Ei}(z) \equiv - \int_{-z}^\infty \frac{e^{-t}}{t} \; dt \qquad 
( | {\rm arg}(z)| < \pi ) \;   ,
$$
in which the integral is defined in the sense of the principal value, with a branch cut 
along the negative real axis \cite{Abramowitz_46}. Then we find
$$
J_q(c) = -\; \frac{3\pi c}{2q^3} \; \left ( q^2 + c^2 \right ) +
\frac{3e^{-c}}{2q^3} \; 
\left [ \left ( 2 - c + c^2 \right ) \sin q - ( 2 - c) q\cos q \right ] +
$$
\be
\label{84}
+
\frac{3ic}{4q^3} \; \left ( q^2 + c^2\right ) 
[ {\rm Ei}( - c - iq) - {\rm Ei}(-c+iq) ]  \;  .
\ee
To show that this expression is real valued, we can employ the series representation for
the exponential integral function
$$
 {\rm Ei}(z) = \gm + \ln z + \sum_{n=1}^\infty \frac{z^n}{n n!} \;  ,
$$
where $\gamma = 0.57721$ is the Euler-Mascheroni constant. Using the relation
$$
 {\rm Ei}(-c-iq) - {\rm Ei}(-c+iq) = 2i(\al-\pi) +
2i\sum_{n=1}^\infty \frac{(q^2+c^2)^{n/2}}{n n!} \;\sin [ n(\al -\pi) ] \; ,
$$
in which 
$$
\al \equiv \arctan \; \frac{q}{c} \;  ,
$$
we obtain
$$
J_q(c) = \frac{3e^{-c}}{2q^3} 
\left [ \left ( 2 - c +c^2 \right ) \sin q - ( 2 - c) q \cos q \right ] -
$$
\be
\label{85}
 -
\; \frac{3c(q^2+c^2)}{2q^3} \left \{ \al + \sum_{n=1}^\infty \frac{(q^2+c^2)^{n/2}}{n n!} \;
\sin [ n (\al-\pi) ] \right \}  \; .
\ee
   
Function (\ref{84}) possesses the following properties. When $q \ra 0$, under a finite $c$,
we have
\be
\label{86}
 J_q(c) \simeq \frac{(1+c)e^{-c}}{5c^2} \; q^2 - \; 
\frac{(6+6c+3c^2+c^3)e^{-c}}{70c^2} \; q^4 \qquad (q\ra 0) \; .
\ee
This defines the long-wave limit of the Fourier transform (\ref{77}) in the form
\be
\label{87}
 D_k(b,\kappa) \simeq D_k \; \frac{(1+\kappa b) e^{-\kappa b}}{5\kappa^2} \; k^2 
\qquad ( k \ra 0) \;  .
\ee
Note that this limit is principally dependent on a finite value of the screening parameter 
$\kappa$ that cannot be set to zero. Thus in the long-wave limit, we get
\be
\label{88}
D_0(b,\kappa) \equiv \lim_{k\ra 0} D_k(b,\kappa) = 0 \;   ,
\ee
that agrees with the integral
\be
\label{89}
 \int D(\br,b,\kappa)\; d\br = 0 \;  ,
\ee
and which is in agreement with condition (\ref{23}).

In the short-wave limit, when $k \ra \infty$, and $c$ being finite, we get
\be
\label{90}
 J_q(c) \simeq - 3e^{-c} \left [ \frac{\cos q}{q^2} \; - \; \frac{1-c}{q^3} \sin q - \;
\frac{c(1+c)}{q^4} \cos q  \right ] \; ,
\ee
which gives for the Fourier transform
\be
\label{91}
 D_k(b,\kappa) \simeq - 3 D_k e^{-\kappa b} \; \frac{\cos(kb)}{(kb)^2} \qquad
( k \ra \infty ) \;  .
\ee
Here the finiteness of the short-range cutoff $b$ is important.   

In the limit of a small screening parameter, under finite $q$, we find
\be
\label{92}
J_k(c) \simeq a_0(q) + a_1(q) c + a_2(q) c^2 \qquad (c\ra 0) \;   ,
\ee
where the coefficient functions are
$$
a_0(q) = \frac{3}{q^3} ( \sin q - q \cos q ) \; ,
$$
$$
a_1(q) = \frac{9}{2q^3}(  q \cos q - \sin q) + 
\frac{1}{2q} \left [ {\rm Si}(q) - \; \frac{\pi}{2} \right ] \; , \qquad
a_2(q) = \frac{3}{2q^3} ( 3 \sin q - q \cos q ) \; ,
$$
and the notation for the sine integral
$$
{\rm Si}(q) = \int_0^q \frac{\sin t}{t} \; dt
$$
is used. When $q\ra 0$, we get
$$
a_0(q) \simeq 1 \; , \qquad a_1(q) \simeq -\;\frac{3\pi}{2q} \; , \qquad
a_3(q) \simeq \frac{3}{q^2} \qquad (q\ra 0) \; .
$$
If we try to find here the long-wave limit, we get the Fourier transform
\be
\label{93}
D_k(b,\kappa) \simeq D_k \left ( 1 - \; \frac{3\pi\kappa}{4k} + 
\frac{3\kappa^2}{k^2} \right ) \qquad ( \kappa \ra 0 , \; k \ra 0 ) \;   .
\ee
This expression is divergent at $k \ra 0$, demonstrating that the expansion in powers
of the screening parameter $(c \ra 0$ is not defined. The function $J_q(c)$ is not 
analytical at $c = 0$, hence, the Fourier transform $D_k(b,\kappa)$ is not analytical
at $\kappa = 0$. 

For a very large screening parameter, we have
\be
\label{94}
 J_q(c) \simeq e^{-c} \left [ \frac{d_1(q)}{c} + \frac{d_2(q)}{c^2} \right ] \qquad
(c\ra \infty) \;  ,
\ee
with 
$$
d_1(q) = \frac{3}{q^3} \left [ \left ( 3 - q^2 \right ) \sin q - 3q\cos q \right ] \; , 
$$
$$
d_2(q) = \frac{3}{q^3} \left [ \left ( 12 - q^2 \right ) q\cos q  + \left ( 5q^2-12
\right ) \sin q \right ] \; .
$$
Here, the long-wave limit of the Fourier transform is defined, yielding
\be
\label{95}
 D_k(b,\kappa) \simeq D_k \; \frac{be^{-\kappa b}}{5\kappa} \; k^2 \qquad
(b\ra\infty, \; k\ra 0 ) \;  .
\ee
   
In this way, in order that the dipolar interaction potential would enjoy a well-defined
Fourier transform, it is necessary to regularize this potential both, for short-range as
well as for long-range interactions. The behavior of the function $J_q(c)$, defining the
Fourier transform of the dipolar potential, is illustrated in Fig. 1.  

The regularized dipolar potential is an effective potential taking into account short-range 
and long-range particle correlations. One may ask whether the use of an effective potential, 
instead of the bare interaction potential, is admissible. The answer is yes. The study of any 
many-particle system can be started with a self-consistent mean-field approximation, containing  
an effective potential, which is called the {\it correlated mean-field approximation} 
\cite{Yukalov_47}. The higher approximations, beyond the correlated mean-field approximation,
can be obtained by means of an iterative procedure for Green functions 
\cite{Yukalov_48,Yukalov_49,Yukalov_50}.

\section{Excitation spectrum}

After the interaction potential is properly regularized, so that it enjoys a well defined 
Fourier transform, it is possible to study the properties of the system. It is necessary
to emphasize that without the regularization the formal investigation of the system 
properties in a mean-field approximation is not correct and would lead to wrong conclusions.
It is admissible to deal with a singular bare interaction potential only in higher-order 
approximations, taking into account particle correlations smearing the singularities in the 
bare potential. However, in a simple mean-field approximation, the use of a bare potential 
having no well defined Fourier transform is inadmissible. But a correlated mean-field 
approximation involving an effective regularized potential is justified.  

Let us consider the excitation spectrum of a Bose-condensed system having the interaction 
potential consisting of two terms, 
\be
\label{96}
\Phi(\br) = 4\pi\; \frac{a_s}{m} \; \dlt(\br) + D(\br,b,\kappa) \;  .
\ee
The first term describes local atomic interactions, with $a_s$ being the scattering length.
And the second term is the regularized dipolar interaction potential. Strictly speaking, the 
scattering length depends on the strength of the dipolar interactions, but in a wide range of
the scattering-length values, it can be varied independently of the dipolar part    
\cite{Griesmaier_14,Baranov_15,Baranov_16}. 

The Fourier transform of potential (\ref{96}) is 
\be
\label{97}
 \Phi_k =  4\pi\; \frac{a_s}{m} + D_k(b,\kappa) \; .
\ee
Keeping in mind that
$$
 \lim_{k\ra 0}  D_k(b,\kappa) = 0 \;  ,
$$
the long-wave limit 
\be
\label{98}
\Phi_0 \equiv \lim_{k\ra 0}  \Phi_k =  4\pi\; \frac{a_s}{m}
\ee
contains only the local interaction part. For arbitrary $k$, the Fourier transform (\ref{97})
can be represented as  
\be
\label{99}
 \Phi_k = \Phi_0 + f_k \;  ,
\ee
where
\be
\label{100}
f_k \equiv D_k (b,\kappa) = D_k I_k(b,\kappa) \;   .
\ee
The dipolar term can be written as
\be
\label{101} 
f_k = D_k J_q(c) \qquad ( q \equiv k b \; , ~ c \equiv \kappa b ) \;    .
\ee
 
For the spectrum of collective excitations  (\ref{50}), we have
\be
\label{102}
 \ep_k^2 = \left [ \frac{k^2}{2m} + (\rho_1 - \sgm_1) f_k \right ]
\left [ \frac{k^2}{2m} + 2 (\rho_0 + \sgm_1) \Phi_0 + 
(2\rho_0 + \rho_1 +\sgm_1)f_k \right ] \;   .
\ee
In the long-wave limit, the spectrum is of phonon type,
\be
\label{103}
\ep_k \simeq c_k k \qquad ( k \ra 0) \;   .
\ee
The sound velocity here is obtained by taking account of the properties of the function 
$$
J_q(c) \simeq A \left ( \frac{k}{\kappa} \right )^2 \qquad ( q\ra 0 ) \;   ,
$$
where
$$
A \equiv \frac{1}{5} ( 1 + \kappa b) e^{-\kappa b} \;   ,
$$
so that
$$
f_k \simeq A D_k \left ( \frac{k}{\kappa} \right )^2   \qquad ( k \ra 0 ) \; .
$$
Thus for the sound velocity, we find 
\be
\label{104}
 c_k^2 = (\rho_0 + \sgm_1) \; \frac{\Phi_0}{m} \; 
\left [ 1 + 2m (\rho_1 - \sgm_1) \; \frac{A}{\kappa^2} \; D_k \right ] \;  .
\ee
The sound velocity is anisotropic because of $D_k$. Although it is necessary to stress that
the anisotropy appears only if $\rho_1$ and $\sigma_1$ are not zero, but are defined by
the expressions
\be
\label{105}
  \rho_1 = \int n_k \; \frac{d\bk}{(2\pi)^3} \; , \qquad 
\sgm_1 = \int \sgm_k \; \frac{d\bk}{(2\pi)^3} \; ,
\ee
in which the functions $n_k$ and $\sigma_k$ are defined in equations (\ref{32}) and (\ref{33}).
The condensate density $\rho_0 = \rho - \rho_1$ is expressed through the density of 
uncondensed atoms $\rho_1$. 

The functions $n_k$ and $\sigma_k$ are connected with each other,
$$
 \sgm_k = -\; \frac{\Dlt_k}{\om_k} \; \left ( n_k + \frac{1}{2} \right ) \;   .
$$
It is important to emphasize that the anomalous average $\sigma_1$ is of order or even 
larger than $\rho_1$. Because of this, they are to be taken into account together or
both omitted. But neglecting the anomalous average $\sigma_1$, while keeping the normal
density $\rho_1$, is principally wrong \cite{Yukalov_12,Yukalov_13,Yukalov_51}.

\section{Bogolubov approximation}

The Bogolubov approximation is applicable for temperature close to zero and asymptotically 
weak interactions, such that both $\rho_1$ and $\sigma_1$ are much smaller than the condensate
density $\rho_0 \ra \rho$. Neglecting $\rho_1$ and $\sigma_1$ in spectrum (\ref{102}) yields
the Bogolubov spectrum
\be
\label{106}
 \ep_k = 
\sqrt{ \frac{\rho}{m} ( \Phi_0 + f_k ) k^2 + \left ( \frac{k^2}{2m} \right )^2 } \;  .
\ee
In the long-wave limit, the spectrum is of phonon type,
\be
\label{107}
 \ep_k \simeq c_B k \qquad ( k\ra 0) \;  ,
\ee
with the sound velocity
\be
\label{108}
 c_B \equiv \sqrt{\frac{\rho}{m}\; \Phi_0 } = \frac{1}{m} \; \sqrt{ 4\pi\rho a_s } \;  .
\ee
As is evident, in the Bogolubov approximation, the sound velocity is isotropic. Anisotropy
arises only in the higher-order approximation, as is seen in the sound velocity (\ref{104}).

But at finite $k$, spectrum (\ref{106}) is anisotropic. We can consider two opposite cases, 
the so-called parallel geometry, with ${\bf k}$ parallel to the dipole direction,
\be
\label{109}
 \bk \cdot \bfe_d = k \qquad (\vartheta_k = 0 ) \;  ,
\ee
so that 
\be
\label{110}
 D_k  = \frac{8\pi}{3} \; d_0^2 \qquad (\vartheta_k = 0 ) \;    ,
\ee
and the perpendicular geometry, when
\be
\label{111}
 \bk \cdot \bfe_d = 0 \qquad \left ( \vartheta_k = \frac{\pi}{2} \right ) \;  ,
\ee 
so that
\be
\label{112}
  D_k  = -\; \frac{4\pi}{3} \; d_0^2 \qquad \left ( \vartheta_k = \frac{\pi}{2} \right ) \;  .
\ee

It is convenient to introduce the dimensionless spectrum
\be
\label{113}
 \ep(q) \equiv \frac{b}{c_B} \; \ep_k \qquad ( q =  k b ) \;  .
\ee
Also, we define the correlation length
\be
\label{114}
 \xi_c \equiv \frac{1}{mc_B} = \frac{1}{\sqrt{4\pi\rho a_s } } \;  .
\ee
Then spectrum (\ref{106}) leads to the expression
\be
\label{115}
  \ep^2(q) = \left [ 1 + \frac{D_q}{\Phi_0} \; J_q(c) \right ] q^2 +
\left ( \frac{\xi_c}{2b} \right )^2 q^4 \; .
\ee
   
Let us denote the spectrum for the parallel geometry as
\be
\label{116}
\ep_\parallel(q) = \ep(q) \qquad  (\vartheta_k = 0 ) \;  ,
\ee
and for the perpendicular geometry, as
\be
\label{117}
\ep_\perp(q) = \ep(q) \qquad  \left ( \vartheta_k = \frac{\pi}{2} \right ) \;  .
\ee

The quantity
\be
\label{118}
a_D \equiv m d_0^2 
\ee
is called the dipolar length. And the ratio 
\be
\label{119}
\al \equiv \frac{4\pi d_0^2}{\Phi_0} = \frac{a_D}{3a_s} 
\ee
characterizes the relative strength of the dipolar interactions with respect to the local
interactions. With these notations, for spectrum (\ref{116}) we get
\be
\label{120}
\ep^2_\parallel(q)  = [ 1 + 2\al J_q(c) ] q^2 + \left ( \frac{\xi_c}{2b} \right )^2 q^4 \; .
\ee
And spectrum (\ref{117}) becomes
\be
\label{121}
\ep^2_\perp(q)  = [ 1 - \al J_q(c) ] q^2 + \left ( \frac{\xi_c}{2b} \right )^2 q^4 \;  .
\ee

Assuming that the screening parameter is inversely proportional to the correlation length,
we have
\be
\label{122}
c \equiv \kappa b = \frac{b}{\xi_c}  \qquad 
\left ( \kappa = \frac{1}{\xi_c} \right ) \;   .
\ee
Then for spectra (\ref{120}) and (\ref{121}), we obtain
\be
\label{123}
\ep_\parallel(q)  = q \; \sqrt{ 1 + 2\al J_q(c) + \frac{q^2}{4c^2} } 
\ee
and, respectively,
\be
\label{124}
\ep_\perp(q)  = q \; \sqrt{ 1 - \al J_q(c) + \frac{q^2}{4c^2} }    .
\ee
One also considers the relative difference between the parallel and perpendicular geometries,
defined by the quantity
\be
\label{125}
\Dlt(q) \equiv 2 \; 
\frac{\ep_\parallel(q)-\ep_\perp(q)}{\ep_\parallel(q)+\ep_\perp(q)} \;  .
\ee

As examples of atoms with magnetic dipoles \cite{Griesmaier_14,Baranov_15,Baranov_16}, it 
is possible to mention $^{52}$Cr, with the magnetic dipole $\mu_0 = 6 \mu_B$ and the dipolar 
length $a_D = 2.4 \times 10^{-7}$ cm, $^{168}$Er, with $\mu_0 = 7 \mu_B$ and 
$a_D = 1.05 \times 10^{-6}$ cm, and $^{164}Dy$, with $\mu_0 = 10 \mu_B$ and 
$a_D = 2.09 \times 10^{-6}$ cm. Some molecules can have the dipolar lengths as large as 
$a_D \sim 10^{-4}$ cm.  
 
The typical behavior of the excitation spectra (\ref{123}) and (\ref{124}) are shown in 
Figs. 2, 3, and 4. Increasing the dipolar interactions leads to the appearance of a roton 
minimum and then to the roton instability. The relative difference (\ref{125}) is shown in 
Figs. 5 and 6. The minimal relative interaction strength $\alpha_{min}$ corresponds to the 
appearance of the roton minimum, which is defined as the occurrence of a zero derivative of 
the spectrum, with respect to the wave vector. The maximal $\alpha_{max}$ is the relative 
dipolar strength at which the roton minimum touches zero, thus displaying the roton 
instability.   

The origin of the roton minimum in systems with dipolar interactions is rather clear, being 
caused by the attractive part of these anisotropic interactions, which results in the
appearance of the negative term in the spectrum (\ref{124}). The possibility of the roton 
minimum in the spectrum of trapped Bose gases with dipolar forces has been theoretically 
considered for quasi-one-dimensional \cite{Giovanazzi_52,Odell_53,Mazets_54,Kurizki_55} and 
quasi-two-dimensional \cite{Santos_56,Macia_57,Fedorov_58} cases. For sufficiently strong 
dipolar forces, Bose gas can become unstable \cite{Goral_59,Goral_60,Lushnikov_61}, which 
also depends on the trap shape. This instability is due to the roton minimum touching zero
\cite{Ronen_62}. 
   
The most intensive experimental studies have been done for $^{52}$Cr, whose Bose-Einstein 
condensation has been experimentally observed \cite{Griesmaier_63,Stuhler_64}. The scattering 
length for these atoms is $a_s = 105 a_B = 0.555 \times 10^{-6}$ cm. With the peak density
$\rho = 3 \times 10^{14}$ cm$^{-3}$, the correlation length becomes 
$\xi_c = 2.186 \times 10^{-5}$ cm. For the dipolar length $a_D = 2.4 \times 10^{-7}$ cm,
the relative dipolar parameter $\alpha = 0.144$. If the short-range cutoff $b$ is of order 
of the scattering length $a_s$, then the screening parameter $c$ is of order $0.025$. 

By employing Feshbach resonance, the scattering length of $^{52}$Cr can be varied in a wide
range, producing pure dipolar trapped gas and reaching its collapse \cite{Lahaye_65,Koch_66}.

The collective spectrum of dipolar chromium gas was measured by Bismut et al. \cite{Bismut_67}. 
Because of the small relative dipolar strength $\alpha = 0.144$, the spectrum does not show
roton minimum. Qualitatively, the measured spectrum \cite{Bismut_67} as well as the relative 
difference $\Delta(q)$ are in agreement with our results for small $\alpha$.

\section{Local-density approximation}

The theory of the previous sections, developed for uniform systems, can be straightforwardly
generalized to trapped atoms, when the trapping potential is sufficiently smooth, such that
the local-density approximation be valid. Then the off-diagonal parts of correlation functions
are treated as fast, while their diagonal parts, as slow. As a result, the spatial dependence, 
induced by the trapping potential, enters only through the atomic densities and density 
distributions that are treated as slow functions of the spatial variables \cite{Yukalov_13}.

In the presence of an external trapping potential $U({\bf r})$, using the local-density 
approximation, we have the following representation for the single-particle density matrix
\be
\label{126}
\rho_1(\br,\br') = \frac{1}{V} \sum_{k\neq 0} n_k(\br) e^{i\bk\cdot(\br-\br')} 
\ee
and for the anomalous average
\be
\label{127}
\sgm_1(\br,\br') = \frac{1}{V} \sum_{k\neq 0} \sgm_k(\br) e^{i\bk\cdot(\br-\br')} \; ,
\ee
where $V$ is a quantization volume, needed only at the intermediate stage, and where
\be
\label{128}
n_k(\br) =\frac{\om_k(\br)}{2\ep_k(\br)} \; 
\coth \left [ \frac{\ep_k(\br)}{2T} \right ] - \; \frac{1}{2} \; , \qquad 
\sgm_k(\br) = -\; \frac{\Dlt_k(\br)}{2\ep_k(\br)} \; 
\coth \left [ \frac{\ep_k(\br)}{2T} \right ] \; .
\ee
Here, we use the notations
\be
\label{129}
 \om_k(\br) = \frac{k^2}{2m} + U(\br) + \rho(\br) \Phi_0 + \rho_0(\br) \Phi_k +
\frac{1}{V} \sum_{p\neq 0} n_p(\br) \Phi_{k+p} \; - \mu_1(\br)   
\ee
and 
\be
\label{130}
\Dlt_k(\br) = \rho_0(\br) \Phi_k + \frac{1}{V} \sum_{p\neq 0} \sgm_p(\br) \Phi_{k+p} \;   ,
\ee
defining the local spectrum of collective excitations
\be
\label{131}
 \ep_k(\br) = \sqrt{\om_k^2(\br) - \Dlt_k^2(\br) } \;  .
\ee

The density of uncondensed atoms and the anomalous average become
$$
\rho_1(\br) = \frac{1}{V} \sum_{k\neq 0} n_k(\br)  = 
\int n_k(\br) \; \frac{d\bk}{(2\pi)^3} \; ,
$$
\be
\label{132}
\sgm_1(\br) = \frac{1}{V} \sum_{k\neq 0} \sgm_k(\br)  = 
\int \sgm_k(\br) \; \frac{d\bk}{(2\pi)^3} \; .
\ee
The total atomic density is
\be
\label{133}
\rho(\br) = \rho_0(\br) + \rho_1(\br) \; , \qquad \rho_0(\br) = |\eta(\br)|^2 \;   .
\ee
The condensate function, defined by condition (\ref{8}), satisfies the equation
$$
\left [ -\; \frac{\nabla^2}{2m} + U(\br) + 
\int \Phi(\br-\br') \rho(\br') \; d\br' \right ] \eta(\br) \; +
$$
\be
\label{134}
+ \int \Phi(\br-\br') [\; \rho_1(\br,\br') + \sgm_1(\br,\br')\; ] \eta(\br') \; d\br' = 
\mu_0 \eta(\br) \; .
\ee
In the Bogolubov approximation, when almost all atoms are condensed, so that the quantities
related to uncondensed atoms and the anomalous average can be neglected, the 
condensate-function equation (\ref{134}) simplifies to the nonlinear Schr\"{o}dinger equation
\be
\label{135}
 \left [ -\; \frac{\nabla^2}{2m} + U(\br) + 
\int \Phi(\br-\br') \rho(\br') \; d\br' \right ] \eta(\br) = \mu_0 \eta(\br) \;   .
\ee
  
The condition of the condensate existence (\ref{41}), which now reads as
\be
\label{136}
\lim_{k\ra 0} \ep_k(\br) = 0 \; ,
\ee
yields
\be
\label{137}
\mu_1(\br) = U(\br) + \rho(\br) \Phi_0 + 
\frac{1}{V} \sum_{k\neq 0} [ n_k(\br) - \sgm_k(\br) ] \Phi_k \;     .
\ee
Substituting this into equation (\ref{129}) gives
\be
\label{138}
 \om_k(\br) = \frac{k^2}{2m} + \rho_0(\br) \Phi_k + 
\frac{1}{V} \sum_{p\neq 0} [ n_p(\br) (\Phi_{k+p}-\Phi_p) + \sgm_p(\br)\Phi_p ] \;  .
\ee
Then for the spectrum of collective excitations (\ref{131}), we have the equation
$$
\ep_k^2(\br) = \left \{ \frac{k^2}{2m} + 
\frac{1}{V} \sum_{p\neq 0} [\; n_p(\br) - \sgm_p(\br) \; ] (\Phi_{k+p}-\Phi_p) \right \} 
\times
$$
\be
\label{139}
\times
\left [  \frac{k^2}{2m} + 2\rho_0(\br) \Phi_k +
 \frac{1}{V} \sum_{p\neq 0} \{ [\; n_p(\br) + \sgm_p(\br)\;] \Phi_{k+p} -
[\; n_p(\br) - \sgm_p(\br) \;] \Phi_{p} \} \right ] \; .
\ee

Resorting to approximation (\ref{45}), we get, instead of (\ref{137}),
\be
\label{140}
 \mu_1(\br) = U(\br) + [\; \rho(\br) + \rho_1(\br) + \sgm_1(\br)\; ] \Phi_0 \;  ,
\ee
instead of (\ref{138}),
\be
\label{141}
\om_k(\br) = \frac{k^2}{2m} + \rho(\br)\Phi_k -  
[ \; \rho_1(\br) - \sgm_1(\br) \; ] \Phi_0 \;   ,
\ee
and for expression (\ref{130}), we have
\be
\label{142}
 \Dlt_k(\br) =  [ \; \rho_0(\br) + \sgm_1(\br) \; ] \Phi_k \;  .
\ee
The equation for the spectrum (\ref{139}) becomes
$$
\ep_k^2(\br) = \left \{ \frac{k^2}{2m} +   
[ \; \rho_1(\br) - \sgm_1(\br) \; ] (\Phi_k - \Phi_0) \right \}
\times
$$
\be
\label{143}
 \times
  \left \{ \frac{k^2}{2m} + [\; \rho(\br) + \rho_0(\br) + \sgm_1(\br)\; ] \Phi_k - 
 [\; \rho_1(\br) - \sgm_1(\br)\; ] \Phi_0 \right \} \; .
\ee
Keeping in mind dipolar interactions, we use the Fourier transform (\ref{99}), which gives
the spectrum equation  
$$
\ep_k^2(\br) = \left \{ \frac{k^2}{2m} +  
 [ \; \rho_1(\br) - \sgm_1(\br)\; ] f_k \right \}   \times
$$
\be
\label{144}
\times
  \left \{ \frac{k^2}{2m} + 2  [ \;\rho_0(\br)  + \sgm_1(\br)\; ] \Phi_0 +
[\; 2\rho_0(\br) + \rho_1(\br) + \sgm_1(\br) \; ] f_k \right \} \; .
\ee

In the Bogolubov approximation, the spectrum equation reduces to the form
\be
\label{145}
 \ep_k^2(\br) = \frac{\rho(\br)}{m} \; (\Phi_0 + f_k ) k^2 +  
\left ( \frac{k^2}{2m} \right )^2 \; .
\ee
In the long-wave limit, the spectrum is phononic,
\be
\label{146}
\ep_k(\br) \simeq c_B(\br) k \qquad ( k \ra 0 ) \;   ,
\ee
with the isotropic sound velocity
\be
\label{147}
  c_B(\br) =\sqrt{ \frac{\rho(\br)}{m} \; \Phi_0 } \; .
\ee

A special case is when the scattering length can be varied by means of the Feshbach 
resonance \cite{Lahaye_65,Koch_66} and even made zero. In an isotropic harmonic trap,
the Bose condensate of $^{52}$Cr becomes unstable and collapses at 
$a_s \approx 15 a_B = 0.794 \times 10^{-7}$ cm. This corresponds to $\alpha \approx 1$. 

When the scattering length is zero, $a_s = 0$, hence $\Phi_0 = 0$, the spectrum in the 
Bogolubov approximation reads as
\be
\label{148}
 \ep^2_k(\br) = \left [ 2\rho(\br) f_k + \frac{k^2}{2m} \right ] \; \frac{k^2}{2m} 
\qquad ( a_s = 0 ) \; .
\ee
In the long-wave limit, using the behaviour of $f_k$ from Sec. 8, we find 
$$
 \ep_k(\br) \simeq \frac{k^2}{2m} \; 
\sqrt{ 1 + \frac{4m}{\kappa^2} \; A \rho(\br) D_k} \qquad (k\ra 0 ) \;  .
$$
Although the roton instability here occurs when the inequality 
$$
\frac{a_D}{\kappa^2}\; \rho(\br) \geq \frac{3}{16\pi A} 
$$
is valid, but such a quadratic spectrum does not satisfy the Landau criterion for 
superfluidity, which means that the condensate is not stable at all. This is in agreement 
with the fact that in a uniform system, Bose condensate with purely dipolar forces is 
not stable. This is also true for a large isotropic trap. But if the trap is of pancake 
shape, the condensate can be stabilized \cite{Lahaye_65,Koch_66}. However, for a strongly 
anisotropic trap, the local-density approximation can be not appropriate.  
   
Recall that at nonzero temperature and finite interactions, the condensate density does not
coincide with the superfluid density, although superfluidity and condensation arise together
at the Bose condensation point. In the self-consistent mean-field approach 
\cite{Yukalov_12,Yukalov_13,Yukalov_17,Yukalov_18,Yukalov_19}, the superfluid 
density writes as 
\be
\label{149}
 \rho_s(\br) = \rho(\br) - \frac{2Q(\br)}{3T} \;  ,
\ee
with the dissipated heat 
$$
Q(\br) = \int  \frac{k^2}{2m}  \; [ n_k(\br) + n_k^2(\br)  - \sgm_k^2(\br) ] \;
\frac{d\bk}{(2\pi)^3} \;  .
$$
Employing the equality
$$
n_k(\br) + n_k^2(\br)  - \sgm_k^2(\br) = \frac{1}{4\sinh^2\{\bt\ep_k(\br)\} } \;   ,
$$
where $\beta = 1/T$, we find
\be
\label{150}
 Q(\br) = \int \frac{k^2}{8m\sinh^2\{\bt\ep_k(\br)\} } \; \frac{d\bk}{(2\pi)^3} \;  .
\ee

\section{Conclusion}

The main message of this paper is that it is necessary to be cautious dealing with nonlocal 
long-range interactions. Such long-range interactions can exist in different finite quantum 
systems \cite{Birman_68}. Before considering Fourier transforms of such long-range interactions,
it is necessary to check whether the considered interaction potential is absolutely integrable.
The absolute integrability is a sufficient condition for the existence of the corresponding 
Fourier transform. If the interaction potential is not absolutely integrable, it may have no
well defined Fourier transform. Then, formally calculating a Fourier transform of a 
nonintegrable potential, one can get an incorrect expression leading to senseless unphysical
results. For instance, one can come to a conclusion that thermodynamic characteristics, such 
as chemical potentials and energies, are not scalars, which certainly has no meaning.

The consideration is specified by the example of a Bose-condensed system with nonlocal 
interactions. The self-consistent mean-field approach is used 
\cite{Yukalov_12,Yukalov_13,Yukalov_17,Yukalov_18,Yukalov_19}. This approach enjoys the unique 
properties, as compared to all other mean-field approximations: It is the sole mean-field theory
guaranteeing the correct second order of the Bose-Einstein condensation transition \cite{Yukalov_69}.
Also, it is the sole mean-field approach yielding the values of the condensate fraction in very 
good agreement with Monte Carlo calculations for arbitrary interaction strength \cite{Yukalov_70}.    

As a particular case, atoms with dipolar interactions are treated. The bare dipolar interaction 
potential, for a three-dimensional system, is not absolutely integrable, and does not have 
a correctly defined Fourier transform. A formally calculated Fourier transform is not well 
defined and leads to wrong conclusions. To get a correctly defined Fourier transform of the
dipolar potential, it is necessary to regularize it, making it absolutely integrable. The
correctly defined Fourier transform of the regularized potential yields the results essentially
differing from the ill-defined transform of the bare potential. Thus, all thermodynamic 
characteristics are scalars, as they should be. The spectrum of collective excitations is
anisotropic. But the sound velocity in the Bogolubov approximation is isotropic. The sound 
velocity becomes anisotropic only in higher approximations, e.g., in the Hartree-Fock-Bogolubov 
approximation taking into account the necessary self-consistency conditions      
\cite{Yukalov_12,Yukalov_13,Yukalov_17,Yukalov_18,Yukalov_19}.

The spectrum of collective excitations and the relative spectrum difference for the parallel
and perpendicular geometries, at a small relative dipolar strength, are in qualitative agreement 
with those experimentally observed \cite{Bismut_67} for the atoms of $^{52}$Cr. However a 
detailed comparison with particular experiments has not been the aim of the present paper. 
This requires a separate publication. The goal of this paper has been to suggest a 
general approach for describing systems with nonlocal long-range interactions and to explain 
the necessity of regularizing the related interaction potentials before taking their Fourier 
transforms.

\vskip 5mm

{\bf Acknowledgement}. Financial support from RFBR (grant $\#$14-02-00723) is appreciated.

\newpage

\begin{center}
{\Large{\bf Figure Captions }}

\end{center}

\vskip 3cm

{\bf Figure 1}. Integral $J_q(c)$, entering the Fourier transform of the regularized dipolar 
interaction potential, as a function of the dimensionless wave vector $q = k b$ for 
different screening parameters $c = \kappa b$, where $b$ is the short-range cutoff.  

\vskip 1cm
{\bf Figure 2}. Spectra of collective excitations in dimensionless units for the parallel 
and perpendicular geometries, as explained in the text, for the relative dipolar 
strength $\alpha = 1$ and the screening parameter $c = 1$. 

\vskip 1cm
{\bf Figure 3}. Spectrum of collective excitations in the perpendicular geometry for the 
screening parameter $c = 0.1$ and different relative strengths of dipolar interactions.
The strength $\alpha_{min}$ corresponds to the appearance of the roton minimum, while
$\alpha_{max}$ shows the dipolar strength, where the roton instability occurs.  

\vskip 1cm
{\bf Figure 4}. Spectrum of collective excitations in the perpendicular geometry for the 
screening parameter $c = 1$ and different relative strengths of dipolar interactions.
The relative dipolar strengths $\alpha_{min}$ and $\alpha_{max}$ show the appearance
of the roton spectrum and the arising roton instability, respectively. 

\vskip 1cm
{\bf Figure 5}. Relative difference between the collective spectra in the parallel and 
perpendicular geometries for the screening parameter $c = 0.1$ and different relative 
dipolar interaction strengths. 

\vskip 1cm
{\bf Figure 6}. Relative difference between the collective spectra in the parallel and 
perpendicular geometries for the screening parameter $c = 1$ and different relative 
dipolar interaction strengths.

\newpage

\begin{figure}[ht]
\centerline{\includegraphics[width=12cm]{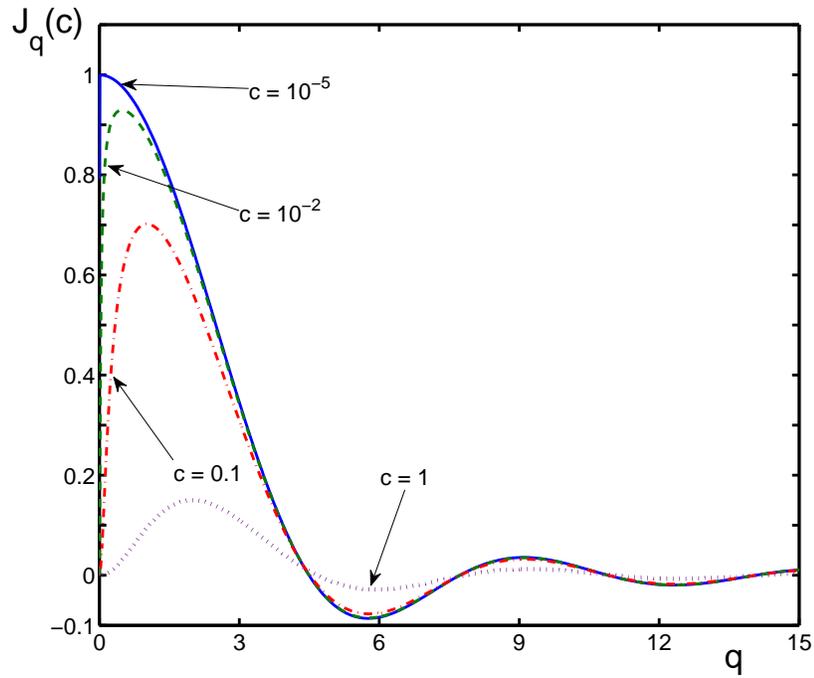} }
\caption{Integral $J_q(c)$, entering the Fourier transform of the regularized dipolar 
interaction potential, as a function of the dimensionless wave vector $q = k b$ for 
different screening parameters $c = \kappa b$, where $b$ is the short-range cutoff.}
\label{fig:Fig.1}
\end{figure}

\begin{figure}[ht]
\centerline{\includegraphics[width=12cm]{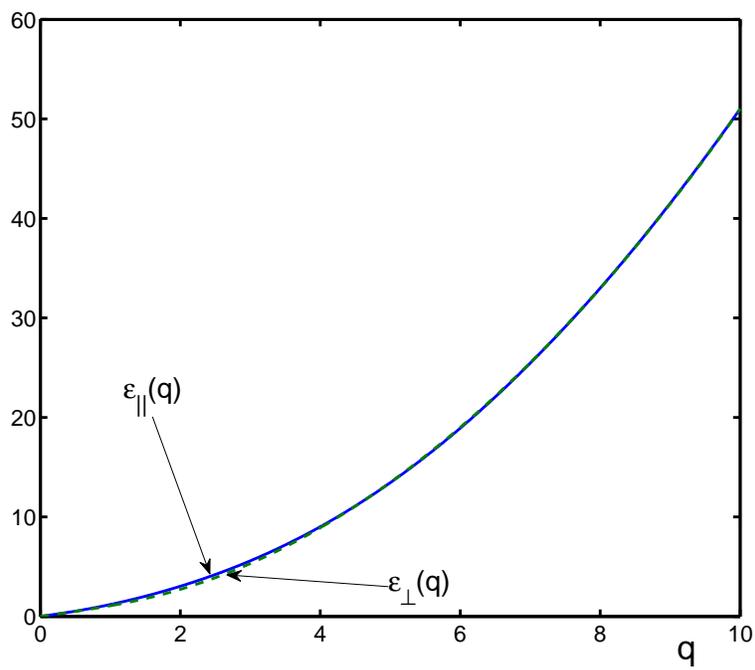} }
\caption{Spectra of collective excitations in dimensionless units for the parallel 
and perpendicular geometries, as explained in the text, for the relative dipolar 
strength $\alpha = 1$ and the screening parameter $c = 1$.}
\label{fig:Fig.2}
\end{figure}

\begin{figure}[ht]
\centerline{\includegraphics[width=12cm]{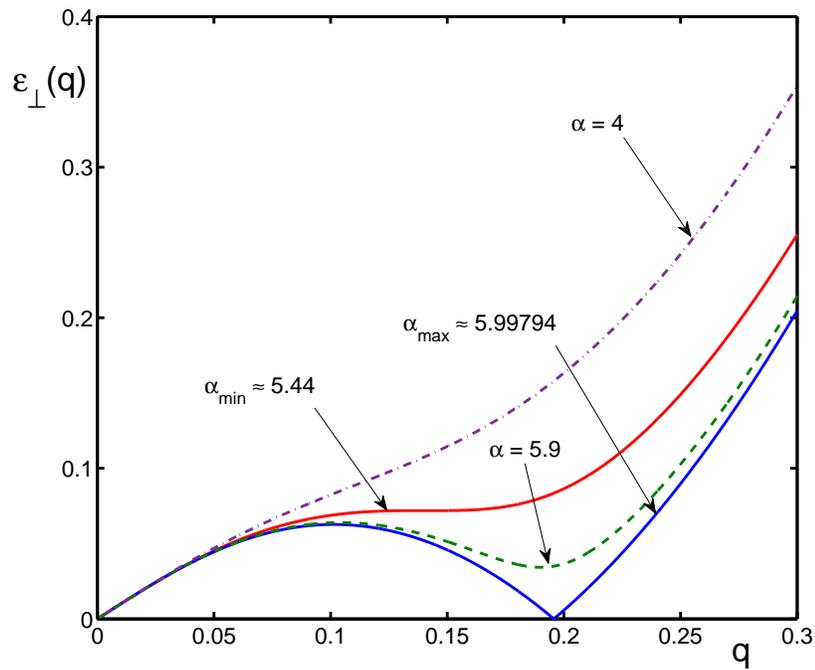} }
\caption{Spectrum of collective excitations in the perpendicular geometry for the 
screening parameter $c = 0.1$ and different relative strengths of dipolar interactions.
The strength $\alpha_{min}$ corresponds to the appearance of the roton minimum, while
$\alpha_{max}$ shows the dipolar strength, where the roton instability occurs. }
\label{fig:Fig.3}
\end{figure}

\begin{figure}[ht]
\centerline{\includegraphics[width=12cm]{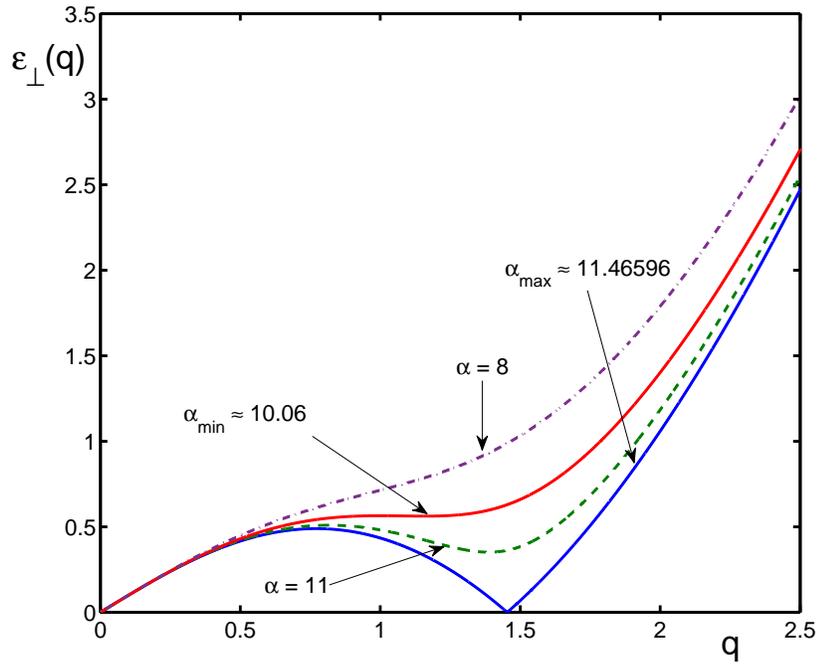} }
\caption{Spectrum of collective excitations in the perpendicular geometry for the 
screening parameter $c = 1$ and different relative strengths of dipolar interactions.
The relative dipolar strengths $\alpha_{min}$ and $\alpha_{max}$ show the appearance
of the roton spectrum and the arising roton instability, respectively.}
\label{fig:Fig.4}
\end{figure}

\begin{figure}[ht]
\centerline{\includegraphics[width=12cm]{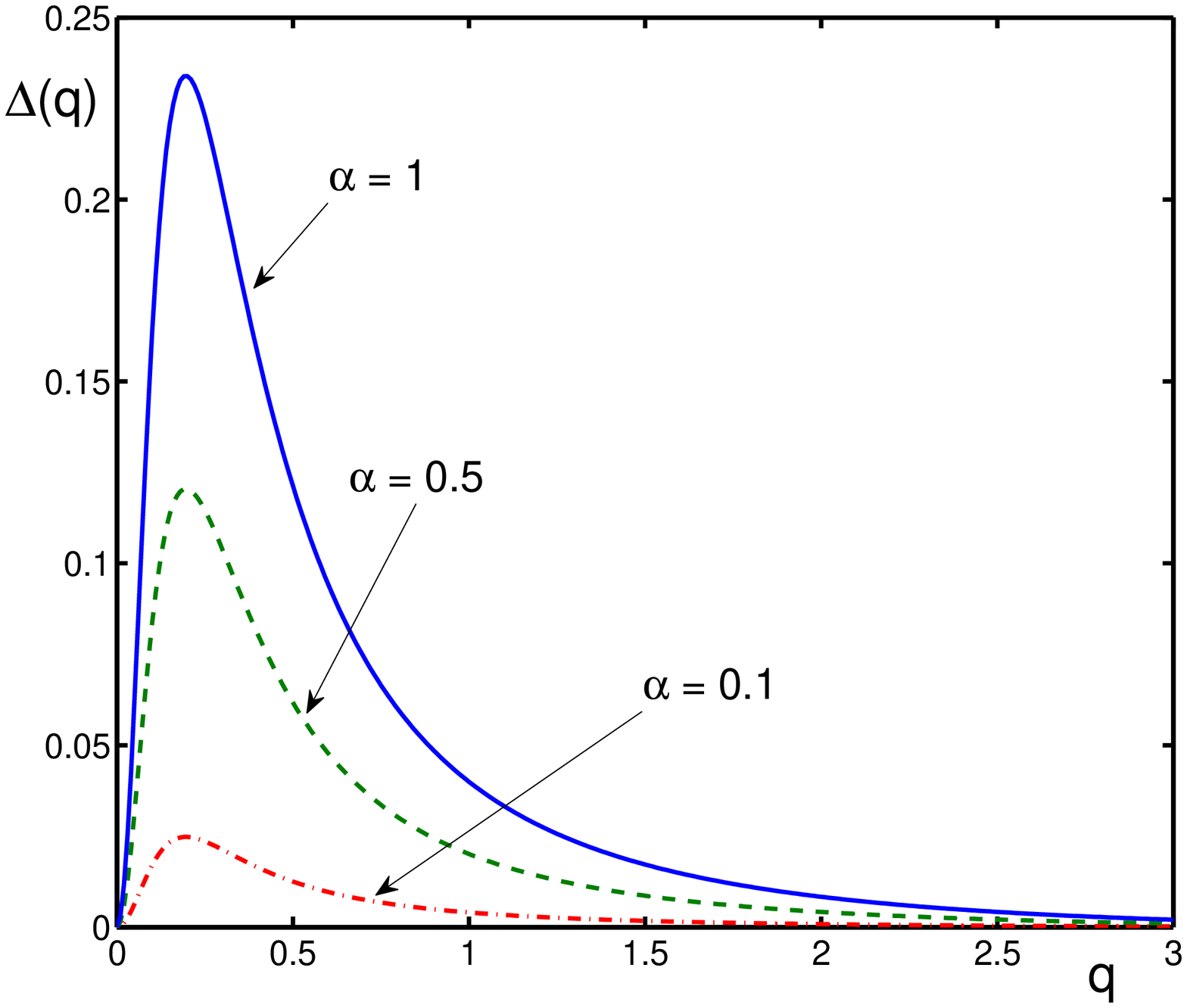} }
\caption{Relative difference between the collective spectra in the parallel and 
perpendicular geometries for the screening parameter $c = 0.1$ and different relative 
dipolar interaction strengths.}
\label{fig:Fig.5}
\end{figure}

\begin{figure}[ht]
\centerline{\includegraphics[width=12cm]{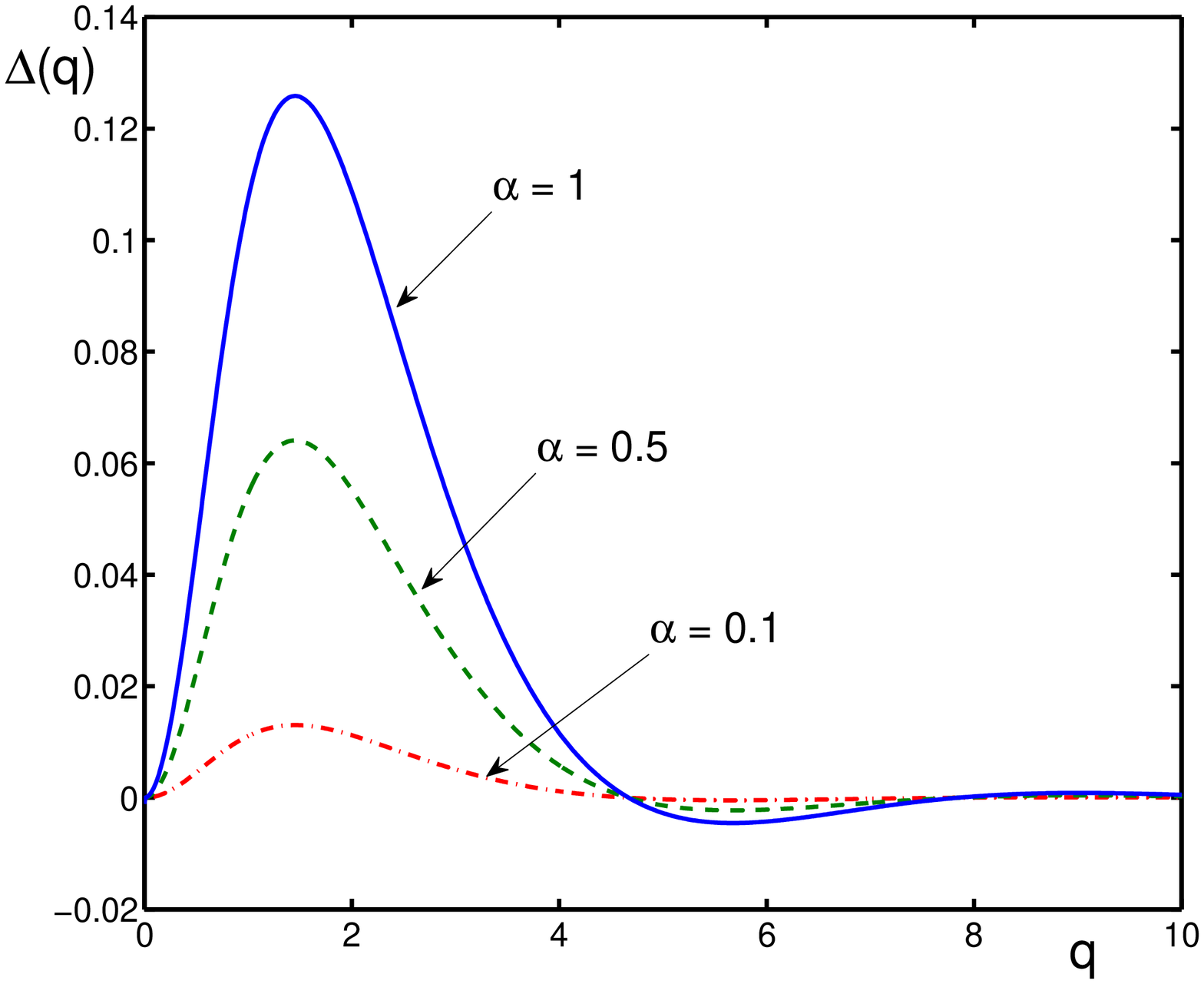} }
\caption{Relative difference between the collective spectra in the parallel and 
perpendicular geometries for the screening parameter $c = 1$ and different relative 
dipolar interaction strengths.}
\label{fig:Fig.6}
\end{figure}

\end{document}